\documentclass[
    ,final            
  ]
  {aipproc}

\layoutstyle{8x11double}

\begin{document}

\title{The X-ray Variability of NGC 4945: Characterizing
the Power Spectrum through Light Curve Simulations}

\author{Martin Mueller}{
  address={Stanford Linear Accelerator Center, 2575 Sand Hill Road, Menlo Park, CA 94025, USA},
}

\author{Greg Madejski}{
  address={Stanford Linear Accelerator Center, 2575 Sand Hill Road, Menlo Park, CA 94025, USA}
}

\author{Christine Done}{
  address={Department of Physics, Durham University, South Rd, DH1 3LE Durham, UK}
}

\author{Piotr Zycki}{
  address={Nicolaus Copernicus Astronomical Center, Bartycka 18, 00-716 Warsaw, Poland}
}

\begin{abstract}
For light curves sampled on an uneven grid of observation times, 
the shape of the power density spectrum (PDS) includes severe 
distortion effects due to the window function, and simulations 
of light curves are indispensable to recover the true PDS.  
We present an improved method for comparing 
light curves generated from a PDS model to the measured data and apply 
it to a 50-day long \emph{RXTE} observations of NGC 4945, a Seyfert 2 
galaxy with well-determined mass from megamaser observations.  
The improvements over previously reported investigations 
include the adjustment of the PDS model normalization for each 
simulated light curve in order to directly investigate how well the 
chosen PDS shape describes the source data. We furthermore implement 
a robust goodness-of-fit measure that does not depend on the form of 
the variable used to describe the power in the periodogram.  
We conclude that a knee-type function (smoothly
broken power law) describes the data better than a simple 
power law;  the best-fit break frequency is $\sim 10^{-6}$ Hz.  

\end{abstract}

\maketitle


\section{Introduction and Observations}

X-ray variability is a prevailing feature of active galactic 
nuclei (AGN). With long data sets from monitoring campaigns 
by X-ray missions such as \emph{Exosat}, \emph{Ginga} and 
especially \emph{RXTE}, it 
became possible to study the shape of the power density 
spectrum (PDS) over many decades of temporal frequency.  Locally, 
the PDS is well-described by a power law, but globally, 
a suppression of power on long time scales
has been found in a number of AGN
\citep{pap95,ede99,chi00,now00,pou01}. 
Because of the direct proportionality
between the mass and the Schwarzschild radius
of a black hole, the timescale at which this suppression
becomes important is expected to obey 
a linear relationship with the mass. Evidence for such
a correlation has indeed started to emerge \citep{mar03,mch03}
and can be extended over several orders of magnitude
to galactic black hole candidates in their low/hard states,
suggesting that the X-ray emission mechanism is similar
for both classes of objects.

Our method for analyzing unevenly sampled light curves 
is based on simulation techniques used by a number of
researchers \citep{mar03,don92,utt02,cze03}, but introduces
significant changes.
We apply it to the Seyfert 2 galaxy NGC 4945, 
which is unique among the AGN for which a break in the
PDS has been detected in that the mass of the central 
black hole of $1.4 \times 10^{6}$ M$_{\odot}$ is known
fairly accurately from 
mapping the H$_{2}$O maser emission \citep{gre97}. 
A precise knowledge of the 
break frequency in this source should thus enable us to 
calibrate the relationship between the break 
frequency and the mass of the black hole. 

\begin{figure}
\label{lc}
  \includegraphics[width=0.47\textwidth]{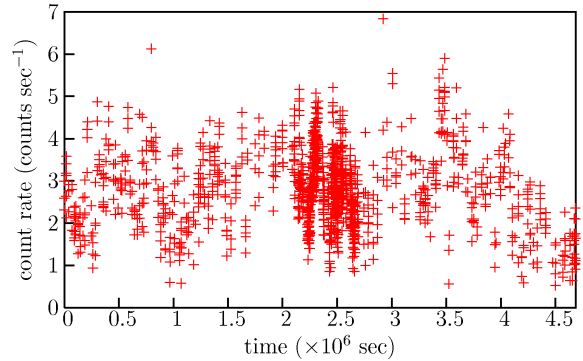}
  \caption{\emph{RXTE} PCA light curve for NGC 4945 in 
the 8--30 keV band obtained in late 2002. Bin size is
variable, with an average of 320 s.
Error bars (not shown) are typically $\pm 0.3$ counts/s.}
\end{figure}

NGC 4945 was observed by \emph{RXTE} in late 2002 for a period spanning a total 
of 50 days. A pointing of 1,400 seconds on average was scheduled 
approximately every 6 hours, while for 7 days centered within the 
total observation period the source was monitored intensively. 
Data from the Proportional Counter Array (PCA) 
were reduced using standard \emph{RXTE} PCA analysis tools.  
The average count rate and the time-averaged photon spectrum are 
consistent with previous observations \citep{don96,mad00}.  
The nuclear flux below 8 keV is 
heavily absorbed at the source, presumably by the same material that 
is responsible for the megamaser activity.  
The counts from all three layers in PCU 0 and 2 
corresponding to the nominal range of energies from 8 to 30 keV were 
added to produce a combined light curve (16 s intrinsic binning), 
resulting in a total of usable data of about 300 ks. The light curve is
shown in figure \ref{lc}.

\section{Data analysis}

\begin{figure}
\label{analysissteps}
  \includegraphics[height=.75\textheight]{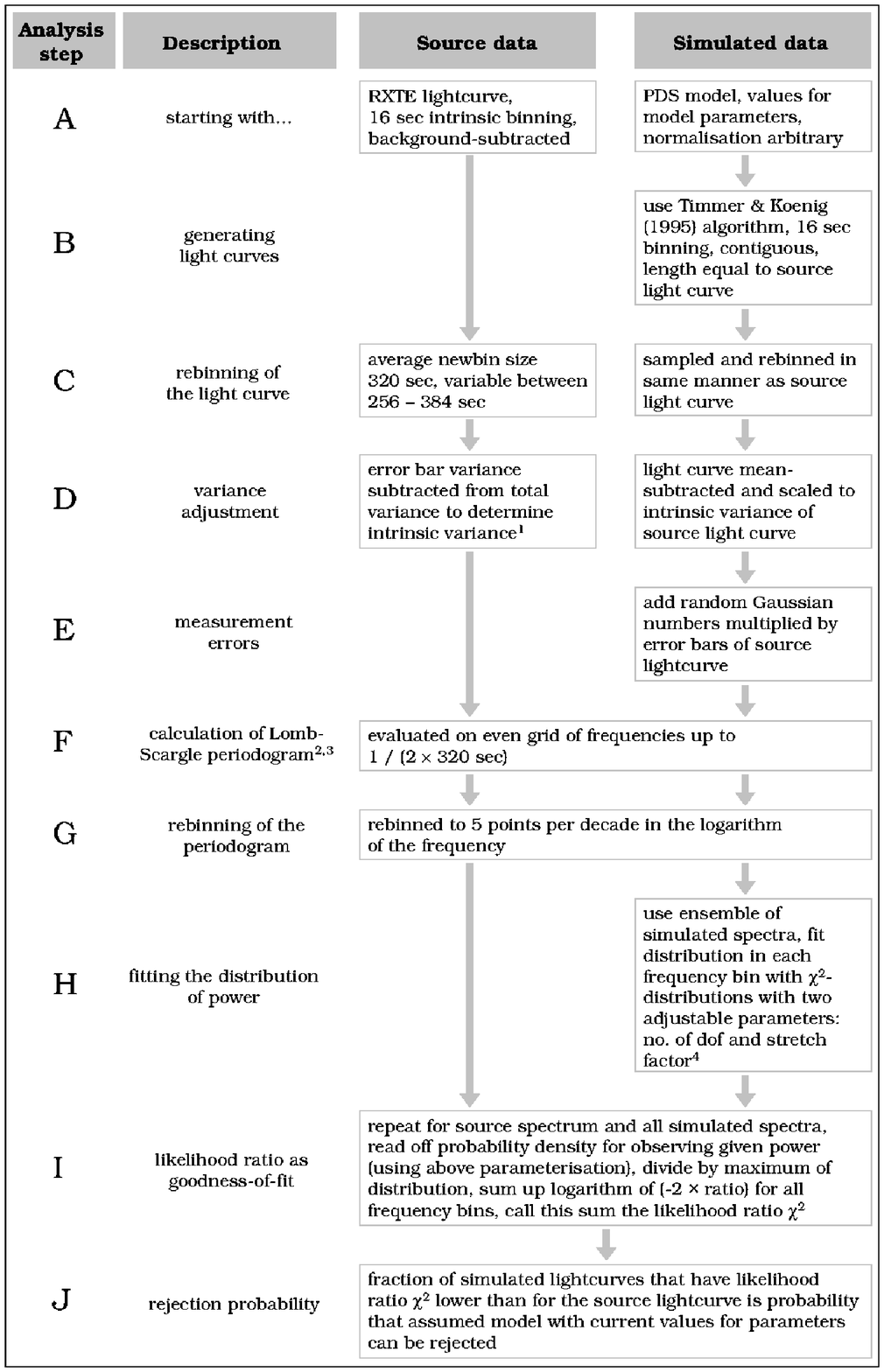}
  \caption{The diagram of the analysis steps used in the 
process of PDS characterization.\newline
\emph{Footnotes to the diagram:}\newline
(1) error bar variance is calculated as
the mean square of the source light curve error bars\newline
(2) see \citet*{lom76} and \citet*{sca82}\newline
(3) Since the rebinned light curve includes contiguous
segments sampled at 320 sec resolution in the intensive section of the
observation, information about the PDS shape can be extracted down to
that timescale.\newline
(4) We are only concerned here with finding
a suitable parameterization of these distributions. $\chi^2$ distributions
with a stretch factor are a natural choice since the power in the 
periodogram is drawn from $\chi^2$ distributions scaled by the power
of the underlying PDS.}
\end{figure}

The algorithm of \citet*{tim95} is used to generate simulated light curves
based on a user-chosen PDS model that are then sampled to match the
observation times of the source light curve. Further processing 
comprises subtracting the mean, scaling each light curve to the 
intrinsic variance of the source light curve, and adding random
Gaussian numbers multiplied by the error bars of the source light
curve. We then calculate the Lomb-Scargle periodogram 
\citep{lom76,sca82}
and re-bin it to 5 points per 
decade.
Using the ensemble of (typically \mbox{$> 500$}) simulated power density 
spectra, we fit the
distribution of the periodogram power in each frequency bin with
a stretched $\chi^2$ distribution: 
$p_{\nu}^{s}(\chi^2,b) = b^{-1} p_{\nu}(\chi^2 / b)$
where $p^s_{\nu}$ is the stretched, $p_{\nu}$ the regular 
$\chi^2$ distribution, $\nu$ is the number of DOF, 
and $b$ is the stretch factor.
Finally, we compare the source periodogram to the fitted distributions, 
using a $\chi^2$-like statistic based on the likelihood ratio as the 
goodness-of-fit estimator \citep{bak84}.  The 
probability that an assumed model with the given set of parameters 
can be rejected is then the fraction of simulated light curves that 
have the likelihood ratio $\chi^{2}$ lower than the source light curve.  
The process is illustrated as a diagram in Fig. \ref{analysissteps}, 
but further 
explanation is outlined below.  

{\bf Scaling of simulated light curves to variance of source data (step D):} 
In previous implementations of light curve simulations, the normalization 
of the PDS model was set to an arbitrary value at first. The best fit 
normalization was then determined by multiplying all simulated spectra 
by the same factor until the fit statistic was minimized 
\citep{mar03,utt02}. 
Due to the 
stochastic nature of the periodogram, the light curves simulated 
from a fixed normalization show a spread in variances \citep{vau03}, 
and the distribution of the power in each frequency 
bin will reflect this spread. By comparing the source data to a 
set of light curves simulated in this way, we are effectively 
asking the question: \emph{How well do the chosen shape and normalization 
of the PDS model describe the variability of the source?}  
In reality, however, we are only interested in characterizing 
the shape of the periodogram. By allowing the normalization to 
vary from one light curve to the next, a slightly different question 
can be investigated: \emph{How likely is it that the source light curve 
was produced by a process that has the chosen PDS shape?}  
By scaling each light curve to the intrinsic variance of the 
source light curve (and adding the Poisson noise level---see below), we 
ensure that the area under the curve for each simulated spectrum 
is equal to the one for the source. 

{\bf Preservation of the Poisson noise level through the normalization 
of the PDS model (step E):}   
The uncertainties in the source count rate are expected to 
contribute a constant level of power in the periodogram (usually called the 
Poisson noise level). Since the raw count rate in the \emph{RXTE} PCA is dominated 
by the background for this source, the errors on the background-subtracted 
light-curve are approximately Gaussian. To mimic the effect of measurement 
errors in the simulated light curves, random Gaussian numbers multiplied by 
the error bars of the source light curve are added. This is done after the 
light curves have been scaled to the correct variance to ensure that the 
Poisson noise level in the simulated light-curves equals the one expected 
in the source data.

{\bf Use of robust fitting statistic to find best-fit parameters 
of the model PDS (steps H--J):}  
A useful goodness-of-fit measure for non-Gaussian distributions is the
rejection probability, calculated as the fraction of simulated light
curves that have a value of the test statistic lower than the 
source data. The choice for test statistic is up to the investigator, 
and an obvious choice is the $\chi^{2}$ statistic, which uses 
the ensemble average and standard deviation of the power in each 
frequency bin for comparison to the source power. Because of the 
non-Gaussian nature of the distribution of power in each frequency 
bin, the applicability of the $\chi^{2}$ statistic is however severely 
limited:  specifically, even small departures 
from Gaussian distributions have a marked effect on the 
standard deviation and thus on the rejection probability.  
Since the distribution of the power is 
more symmetric when plotted in the logarithm compared to 
the linear power, \emph{substantially different goodness-of-fit 
values can be obtained simply by a change of variable.}  
The rejection probability, in effect, not only measures how 
well the model (and its associated parameters) describes 
the source data, but also the degree of non-Gaussianity.  
In contrast, by fitting the distributions of the power in each 
frequency bin with suitably parameterized functions (step H in 
the analysis), the effect of variable transformation on the 
fitting statistic is removed: If stretched $\chi^{2}$ distributions 
are an adequate description of the distribution of linear 
power, then under a change of variable the data can equally well 
be fit with the similarly transformed functions. The probability 
density for observing the source power as well as the value at 
the peak of the distribution is unchanged under these 
transformations. \emph{The likelihood ratio is thus an invariant 
quantity,} i.e. independent of the chosen form of the variable
used to describe the power in the periodogram.

\section{Results}

\begin{figure}
\label{unbroken}
  \includegraphics[width=0.47\textwidth]{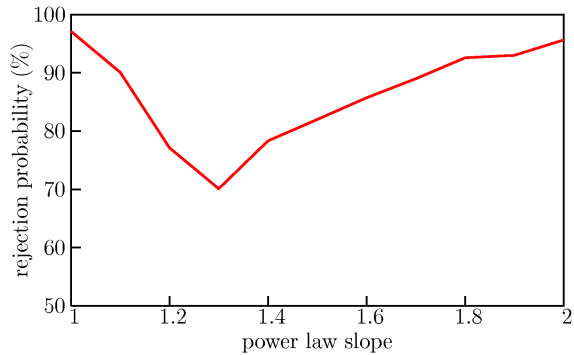}
  \caption{Rejection probability for the unbroken power law model,  The
best fit slope is 1.3, with a rejection probability of 70\% indicating 
a poor fit.}
\end{figure}

\begin{figure}
\label{uttleyknee}
  \includegraphics[width=0.47\textwidth]{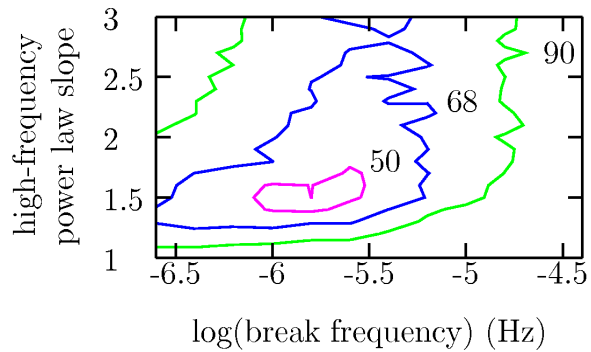}
  \caption{Rejection probability for the knee model, showing contours
for 50, 68, and 90\%.
The best fit is obtained with a break frequency $f_b = 10^{-6}$ Hz 
and a slope 
$\alpha = 1.5$.  The corresponding rejection probability 
is 40\%.    }
\end{figure}

We investigated two models: the simple (unbroken) power law (power law
slope as only parameter), and the knee model of \citet*{utt02} 
(\mbox{$p(f) = A [1 + (f / f_b)^2]^{-\alpha / 2}$;} parameters: high frequency
power law slope $\alpha$, break frequency $f_b$). Figures \ref{unbroken} and
\ref{uttleyknee} show the rejection probabilities obtained. The unbroken
power law results in a minimum rejection probability of 70\% at a slope of
1.3, indicating a rather poor fit. The rejection probability is reduced to
40\% by introducing a $10^{-6}$ Hz break at a high frequency slope of 1.5.
It is clear that the exact location of the break is dependent on the
particular PDS model used to characterize the source variability.

{\bf Uncertainty on rejection probability:}  
To quantify the effect of simulating a finite number of 
light curves from which the goodness-of-fit statistic is 
estimated, we simulated more than 800,000 
light curves using the unbroken power law model with a slope of 1.3.  
The process of fitting the distributions of power through calculating 
the rejection probability (steps H through J in Figure \ref{analysissteps}) 
was repeated 
for sets of light curves of increasing size. The mean 
of the distribution of rejection probabilities is consistent 
across the different sizes, while the standard deviation 
decreases from about 3\% to 2.5\% by going from 500 light curves 
to over 16,000. Assuming that the spread of rejection probabilities 
is only a weak function of the model and its associated parameters, 
we quote a characteristic uncertainty on our fitting statistic of 
about $\pm 3$\%, based on 500 simulated light curves.

{\bf Simple power law vs. knee model:}  The difference in rejection 
probability between the unbroken power law and the knee model is much 
larger than the above uncertainty. The flattening of the power towards low 
frequencies in NGC 4945 has thus been detected at a statistically 
significant level. The length of our observation is most likely not sufficient 
to distinguish between models that use different functional forms for 
the break in the PDS. In particular, the differences in rejection 
probabilities between the knee model and the broken power law model used 
in \citet*{utt02} is not expected to be statistically 
significant for the currently available light curve of NGC 4945.

\end{document}